\documentclass{article}
\usepackage{spconf,amsmath,graphicx}
\usepackage[dvipsnames]{xcolor}
\usepackage{booktabs}       
 \usepackage{amssymb}

\usepackage{url}
 \usepackage{breqn}

\title{SOUNDLOCD: AN EFFICIENT CONDITIONAL DISCRETE CONTRASTIVE LATENT DIFFUSION MODEL FOR TEXT-TO-SOUND GENERATION}
\name{Xinlei Niu$^1$, Jing Zhang$^1$, Christian Walder$^{2}$, Charles Patrick Martin$^1$} 
\address{$^1$Australian National University, Canberra, Australia; $^2$Google DeepMind, Montreal, Canada}
\begin{document}
%
\maketitle
\begin{abstract}
We present SoundLoCD, a novel text-to-sound generation framework, which incorporates a LoRA-based conditional discrete contrastive latent diffusion model. Unlike recent large-scale sound generation models, our model can be efficiently trained under limited computational resources. The integration of a contrastive learning strategy further enhances the connection between text conditions and the generated outputs, resulting in coherent and high-fidelity performance. Our experiments demonstrate that SoundLoCD outperforms the baseline with greatly reduced computational resources. A comprehensive ablation study further validates the contribution of each component within SoundLoCD\footnote{Demo page: \url{https://XinleiNIU.github.io/demo-SoundLoCD/}}.

\end{abstract}
\section{Introduction}
\label{sec:intro}

Sound synthesis spans a diverse array of applications, including game development, film post-production, virtual reality, and musical performance. In these domains, automated sound effects generation could address tasks such as replicating environmental ambience and simulating a specific physical event where the generated sound is required to be aligned with predefined scene descriptions. The challenge of synthesizing environmental sound effects based on a natural language text description is then referred to as \textit{text-to-sound} (T2S) generation.

The pioneering T2S work DiffSound~\cite{diffsound} introduced the concept of using natural language text descriptions as conditions for generating acoustic scenes. Diffsound follows a conditional discrete diffusion pipeline~\cite{GuCBWZCYG22} which involves a pre-trained spectrogram VQ-VAE, the text encoder of a contrastive language-image pretraining model~\cite{CLIP}, and a discrete latent diffusion model (LDM). 
Following~\cite{diffsound}, AudioGen~\cite{AudioGen} proposed a GAN-based auto-regressive model for audio waveform generation under a neural audio codec technique~\cite{ZeghidourLOST22,LiTRUR21}
, where they used the text encoder of a text-to-text transformer~\cite{T5}. However, the higher inference time of AudioGen makes it unsuitable for modelling long sequences.
%
Subsequent studies~\cite{AudioLDM,LLM,Text-Driven} have employed large LDMs~\cite{GuCBWZCYG22,Scorebased,DPM} for generating audio using text prompts in the continuous latent space of a VAE.
Among them, AudioLDM~\cite{AudioLDM} worked on text-free audio generation. \cite{Text-Driven,LLM} focused on tuning a powerful text encoder based on AudioLDM. These prompt models require large-scale data for training. Meanwhile, \cite{makeaudio,make-audio2} proposed the generalization for X-to-Audio generation.  In contrast to those works, we follow DiffSound and focus on a text-to-sound effects LDM model under a discrete latent space of VQ-VAE~\cite{vqvae}. As language is discrete, it is arguably more suitable to integrate textual features into a discrete latent space, facilitated by an LDM.

We observe two main limitations of our baseline model, namely Diffsound~\cite{diffsound}. Firstly, there exist no explicit constraints to achieve effective controllable generation given text conditions. Secondly, the significant requirement of computational resources (see Table~\ref{tab:overal}) poses challenges for other downstream tasks. For the former, most existing conditional LDM models~\cite{diffsound,AudioLDM,LLM,makeaudio,make-audio2} learn the connection between conditions and outputs by adding a conditional prior on the variational lower bound~\cite{GuCBWZCYG22,DhariwalN21}.
We note that this conditional prior does not invariably guarantee a strong linkage between outputs and conditions throughout each diffusion process. This may critically affect the accuracy of generated sounds. For the latter, we claim that an efficient T2S model is more desirable to scale to other related fields.

\begin{figure*}
    \centering
    \vskip -0.1in
    \includegraphics[width = 0.67\textwidth]{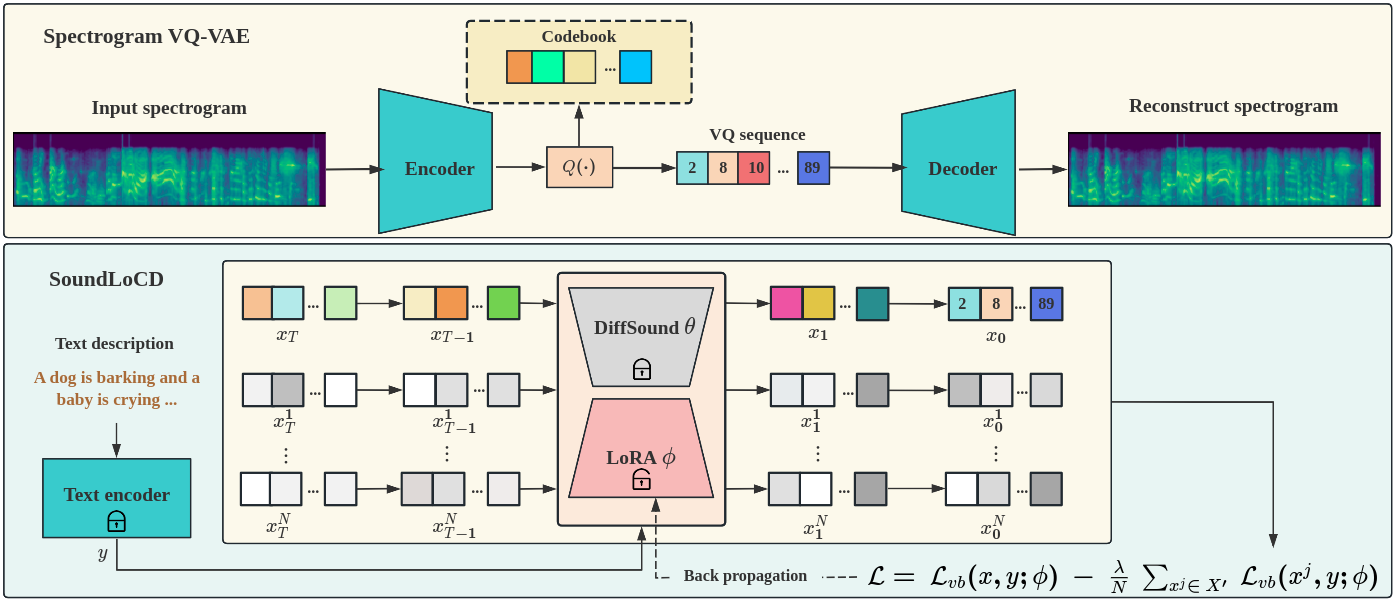}
    \caption{Overall pipeline of SoundLoCD, which is performed based on a pre-trained spectrogram VQ-VAE.
    SoundLoCD involves $N+1$ parallel discrete diffusion processes on original data and $N$ randomly shuffled negative data.
    }
    \label{fig:overall}
    \vskip -0.15in
\end{figure*}
In this work, we propose a novel T2S framework, SoundLoCD, that achieves effective training on small computational devices while further enhancing the connection between text conditions and synthesized sounds. SoundLoCD is a conditional discrete contrastive diffusion model~\cite{CDCD} that injects a smaller amount of trainable parameters on a pre-trained DiffSound~\cite{diffsound} transformer by LoRA~\cite{LoRA}.
Our method obtains a better performance in generating outputs that closely correspond to text conditions while also markedly enhancing training efficiency under limited computational resources.

\section{Methodology}
The overall pipeline of SoundLoCD is in Figure~\ref{fig:overall}, where it learns the latent distribution of the spectrogram VQ-VAE given text conditions with a \textit{contrastive latent diffusion model}. Specifically, given a text-spectrogram pair, we first obtain the VQ sequence $x_0 \in \mathbb{Z}^D$ of the spectrogram with a pre-trained VQ-VAE (same as the VQ-VAE in~\cite{diffsound}). The encoder encodes the spectrogram into a VQ sequence $x_0$ during training, where $D$ is the length of $x_0$. 
The $i^{th}$ token in $x_0$ takes the index that specifies the VQ codebook entries with size $K$. 
The decoder is used to reconstruct the generated VQ sequence $x_0$ of SoundLoCD back to a spectrogram during inference.
Meanwhile, text descriptions are encoded by a pre-trained text encoder to extract the text feature defined as $y\in \mathbb{Z}^M$.
\subsection{SoundLoCD}
SoundLoCD is a LoRA-based \textit{conditional discrete contrastive diffusion}~\cite{CDCD} (CDCD) model which contains two sets of parameters $\{\theta,\phi\}$. $\theta$ is the set of parameters in the DiffSound transformer, which is frozen during training, and $\phi$ is the set of LoRA parameters to be learned. 
As a diffusion model, the forward process corrupts $x_0$ to a pure noise $x_T$ by fixed $T$ time steps, and the reverse process gradually denoises the latent variable to $x_0$ by sampling from $q(x_{t-1}|x_t,x_0)$~\cite{song2020denoising}. Thus, SoundLoCD is trained to approximate the conditional transit distribution $p_{\{\theta,\phi\}}(x_{t-1}|x_t,y)$.

In order to further enhance the connection between $y$ and generated VQ sequence $x_0$, the SoundLoCD involves a contrastive learning strategy and performs a \textit{parallel diffusion process} on a set of negative data $X' = \{x^1,x^2,...,x^N\}$, which contains $N$ randomly shuffled VQ sequences given $x_0$. 
As in~\cite{CDCD}, the idea of involving $X'$ is to pull the well-aligned conditions and the VQ sequence closer, while pushing other pairs apart.
Thus, with contrastive learning, the SoundLoCD aims to produce sound effects that correspond accurately with the condition of text description feature $y$.

\subsubsection{LoRA}
We notice that the performance of existing models relies on huge computational resources
, making it hard to further edit the models. In SoundLoCD, LoRA is applied to adapt a pre-trained DiffSound transformer by injecting a smaller amount of trainable parameters $\phi$ in the transformer, leading to an efficient training system. 
Given a pre-trained DiffSound transformer with a weight matrix $W_0 \in \mathbb{R}^{d \times k}$, LoRA constrains its update by representing the latter with a \textit{low-rank decomposition} $W_0 +\Delta W = W_0 +BA$ where $B \in \mathbb{R}^{d \times r}, A \in \mathbb{R}^{r \times k}$, and the rank $r \ll min(d,k)$. 
Thus the forward pass of SoundLoCD becomes:
\begin{equation}
    h = W_0x +\Delta Wx = W_0 x + BAx,
\end{equation}
where $\{A,B\}\in \phi$. At the beginning of training, $A, B$ are initialized by a Gaussian distribution and zeros respectively. Then $\Delta Wx$ is scaled by $\frac{\alpha}{r}$, $\alpha$ is a constant and $r$ is the rank size.
\begin{table*}[t]
  \renewcommand{\arraystretch}{0.85}
  \renewcommand{\tabcolsep}{3.2mm}
  \vskip -0.2in
  \caption{Model comparison on AudioCaps dataset. DiffSound$\star$ was obtained from the released official checkpoint on AudioCaps. DiffSound$\ast$ is our reproduced result using the original code on AudioCaps. T5-S stands for the T5-small text encoder.}
  \label{tab:overal}
  \centering
  \small
  \begin{tabular}{cccccccc}
    \toprule
    Model   &  Train Config. & Params. & FID $\downarrow$ & ISc $\uparrow$ & KL $\downarrow$   & KID $\downarrow$ \\ 
    \midrule
    DiffSound & 16 Nvidia V100 (2 days)&  434.23M & 13.47 & -  & 4.95 & - \\
    DiffSound$\star$ & 16 Nvidia V100 (2 days) &  434.23M & 12.45 & 16.95 $\pm$ 0.98 &  \textbf{3.85} & 0.00233 $\pm$ 0.00025 \\
    DiffSound$\ast$ & 2 Nvidia RTX3090 (14 days) & 434.23M & 14.96 & 15.87 $\pm$ 1.15 & 4.39 & 0.00424 $\pm$ 0.00022\\
    SoundLoCD + T5-S  & 2 Nvidia RTX3090 (3 days)  & 2.38M & 23.99 & 9.84 $\pm$ 0.39 & 5.23 & 0.00868 $\pm$ 0.00041\\
    SoundLoCD + CLAP  & 2 Nvidia RTX3090 (3 days)  & 2.38M &  36.28 & 5.12 $\pm$ 0.42 & 6.77 & 0.01903 $\pm$ 0.00048 \\
    SoundLoCD + CLIP & 2 Nvidia RTX3090 (3 days)  & 2.38M & \textbf{12.27} & \textbf{17.36 $\pm$ 1.74}  & \textbf{3.85} & \textbf{0.00212 $\pm$ 0.00021} \\
     \bottomrule
  \end{tabular}
  \vskip -0.15in
\end{table*}
\subsubsection{Conditional Discrete Contrastive Diffusion }
SoundLoCD involves $N+1$ parallel discrete diffusion processes for each of the samples in original data $x_0$ and $X'$. 
\begin{table}
\renewcommand{\arraystretch}{0.8}
\renewcommand{\tabcolsep}{1.7mm}
\vskip -0.2in
\caption{Model comparison for fine-tuning on ESC50}
\begin{center}
\begin{small}
\begin{tabular}{cccccc} 
 \toprule
  & FID $\downarrow$  & ISc $\uparrow$  & KL $\downarrow$ & KID $\downarrow$\\ 
 \midrule
 DiffSound & 6.47 & 11.27 $\pm$ 1.21 & 2.54 & 0.00090 $\pm$ 0.00008 \\
 SoundLoCD & \textbf{6.18} & \textbf{10.85 $\pm$ 0.92} & \textbf{2.37} & \textbf{0.00075 $\pm$ 0.00007} \\
  \hline
\end{tabular}
\vskip -0.3in
\label{tab:esc50}
\end{small}
\end{center}
\end{table}

\noindent\textbf{Discrete diffusion process.} To enhance clarity for readers, we present a detailed discrete diffusion process by excluding negative data $x^j \in X'$.
The forward discrete diffusion process gradually corrupts the $x_0$ via a fixed Markov chain $q(x_t|x_{t-1})$ in a fixed number of timesteps $T$.
\begin{equation}
    q(x_t|x_{t-1}) = \mathbf{v}^T(x_t)\mathbf{Q}_t \mathbf{v}(x_{t-1}),
    \label{eq:forward}
\end{equation}
where $\mathbf{v}(x)$ is an one-hot column vector whose length is $K$ and only the entry $x$ is $1$. $\mathbf{Q}$ is a transition matrix $[\mathbf{Q}_t]_{mn} = q(x_t = m|x_{t-1} = n) \in \mathbb{R}^{K\times K}$. The categorical distribution over $x_t$ is given by the vector $\mathbf{Q}_t \mathbf{v}(x_{t-1})$.

Since the Markov chain can marginalize out the intermediate steps and derive the probability of $x_t$ at arbitrary timestep directly from $x_0$ as follows:
\begin{equation}
    q(x_t|x_0) = \mathbf{v}^T(x_t)\overline{\mathbf{Q}}_t \mathbf{v}(x_0), \text{ with } \overline{\mathbf{Q}}_t = \mathbf{Q}_t \cdots \mathbf{Q}_1 \label{eq:margin} 
\end{equation}
The non-Markovian posterior $q(x_{t-1}|x_t,x_0)$ of the diffusion process can be computed according to Equation~\ref{eq:forward} and Equation~\ref{eq:margin}, as
\begin{equation}
    q(x_{t-1}|x_t,x_0) = \frac{(\mathbf{v}^T(x_t)\mathbf{Q}_t \mathbf{v}(x_{t-1})) (\mathbf{v}^T(x_{t-1})\overline{\mathbf{Q}}_{t-1} \mathbf{v}(x_0))}{\mathbf{v}^T(x_t)\overline{\mathbf{Q}}_{t} \mathbf{v}(x_0)}.
    \label{eq:posterior} \nonumber
\end{equation}
\noindent\textbf{Mask-and-replace diffusion strategy.}
Following~\cite{GuCBWZCYG22}, the transition matrix $\mathbf{Q}_t\in \mathbb{R}^{(K+1)\times (K+1)}$ is defined as
\begin{equation}
    \mathbf{Q}_t = 
    \begin{bmatrix}
    \alpha_t + \beta_t & \beta_t & \beta_t & \cdots & 0 \\
    \beta_t & \alpha_t +  \beta_t & \beta_t & \cdots & 0 \\
    \beta_t &  \beta_t & \alpha_t + \beta_t & \cdots & 0 \\
    \vdots &  \vdots & \vdots  & \ddots  & \vdots  \\
    \gamma_t & \gamma_t & \gamma_t  & \cdots & 1 \\
    \end{bmatrix},
\end{equation}
where each ordinary token has a probability of $\gamma_t$ to be a [MASK] token and has a chance of $K\beta_t$ to be uniformly diffused, leaving a probability of $\alpha_t = 1-K\beta_t -\gamma_t$ to be unchanged. The [MASK] token always keeps its own state.
Thus $q(x_t|x_0)$ can be calculated according to 
\begin{equation}
    \overline{\mathbf{Q}}_t \mathbf{v}(x_0) = \overline{\alpha}_t \mathbf{v}(x_0) + (\overline{\gamma}_t - \overline{\beta}_t)\mathbf{v}(K+1) +\overline{\beta}_t,
\end{equation}
where $\overline{\alpha}_t = \prod^t_{i=1} \alpha_i, \overline{\gamma}_t = 1- \prod^t_{i=1}(1-\gamma_i)$, and $\overline{\beta}_t = (1-\overline{\alpha}_t - \overline{\gamma}_t)/K$. And the prior $p(x_T)$ is defined as 
\begin{equation}
    p(x_T) = [\overline{\beta}_T, \overline{\beta}_T, \cdots, \overline{\beta}_T, \overline{\gamma}_T]^T.
    \label{eq:prior}
\end{equation}
\noindent\textbf{Learning.}
To train SoundLoCD, the DiffSound parameters $\theta$ are frozen and only LoRA parameters $\phi$ have been optimized. The overall loss of SoundLoCD is defined as
\begin{equation}
    \mathcal{L} = \mathcal{L}_{vb}(x,y;\phi) - \frac{\lambda}{N} \sum_{x^j\in X'} \mathcal{L}_{vb}(x^j,y;\phi) ,\label{eq:loss}
\end{equation}
where $N$ is the number of negative shuffled VQ sequences in $X'$ and $\lambda$ is the contrastive loss weight. $\mathcal{L}_{vb}(x,y)$ is a conditional variant of the diffusion loss function~\cite{GuCBWZCYG22,CDCD},
\begin{align}
    &\mathcal{L}_{vb}(x,y; \phi) = \mathbb{E}_{q(x_0)}[D_{KL} [q(x_T|x_0)||p(x_T)] + \\
    ~& \sum^T_{t=1} \mathbb{E}_{q(x_t|x_0)}[D_{KL}(q(x_{t-1}|x_t,x_0)||p_{\{\theta,\phi\}}(x_{t-1}|x_t,y))]\nonumber \\
     ~ & = \mathcal{L}_0 + \mathcal{L}_1 + \cdots + \mathcal{L}_{T-1} + \mathcal{L}_T,
\end{align}
where
\begin{equation}
\begin{aligned}
    \mathcal{L}_0 & = -\text{log} p_{\{\theta,\phi\}}(x_0|x_1,y)\\
   \mathcal{L}_{t-1} & = D_{KL}(q(x_{t-1}|x_t,x_0)|| p_{\{\theta,\phi\}}(x_{t-1}|x_t,y))\\
   \mathcal{L}_T &  = D_{KL}(q(x_T||x_0)||p(x_T))
\end{aligned}
\end{equation}

\begin{table*}
  \renewcommand{\arraystretch}{0.8}
  \renewcommand{\tabcolsep}{4.5mm}
\vskip -0.2in
\caption{Performance comparison of different LoRA implementation methods in SoundLoCD with CLIP.} 
\begin{center}
\begin{small}
\begin{tabular}{ccccccc} 
 \toprule
 LoRA config. & Rank ($r$)  & Params. & FID $\downarrow$ & ISc $\uparrow$ & KL $\downarrow$   & KID $\downarrow$  \\ 
  \midrule
  $W_q$ \& $W_k$ & 4 & 583K & 13.70 & 17.12 $\pm$ 1.59 & 3.91 & 0.00231 $\pm$ 0.00023 \\ 
  $W_q$ \& $W_k$ \& $W_v$  & 4  & 836K & 12.54 & 17.16 $\pm$ 1.31 & 4.06 & 0.00226 $\pm$ 0.00021 \\
  $W_q$ \& $W_k$ \& $W_v$ \& $W_p$  & 4  & 1.11M & 13.31 & 16.84 $\pm$ 1.19 & 3.94 & 0.00270 $\pm$ 0.00022\\ 
  \midrule
  $W_q$ \& $W_k$  & 8  & 1.11M & 12.51 & 16.41 $\pm$ 1.79 & 3.88 & 0.00243 $\pm$ 0.00025 \\ 
  $W_q$ \& $W_k$ \& $W_v$ & 8  & 1.63M & \textbf{12.15}  & 17.30 $\pm$ 0.57 &  4.00&  0.00218 $\pm$ 0.00021\\ 
  $W_q$ \& $W_k$ \& $W_v$ \& $W_p$  & 8 & 2.38M & \underline{12.27} & \textbf{17.36 $\pm$ 1.74}  &  \textbf{3.85} & \underline{0.00212 $\pm$ 0.00021}\\ 
  \midrule
  $W_q$ \& $W_k$ & 16  &  2.23M & 12.85 & 16.78 $\pm$ 1.51 &  3.85 &  0.00247 $\pm$ 0.00023\\ 
  $W_q$ \& $W_k$ \& $W_v$  & 16  & 3.27M &12.87 & 16.91 $\pm$ 1.42&  4.01 &  0.00240 $\pm$ 0.00024\\ 
  $W_q$ \& $W_k$ \& $W_v$ \& $W_p$ & 16 & 4.45M & 12.48 & 17.32$\pm$ 1.78  &  3.90 &  \textbf{0.00209 $\pm$ 0.00019}\\ 
 \hline
\end{tabular}
\vskip -0.25in
\label{tab:LORA_cofg}
\end{small}
\end{center}
\end{table*}

\begin{figure}
    \vskip -0.1in
    \centering
    \includegraphics[width = 0.47\textwidth]{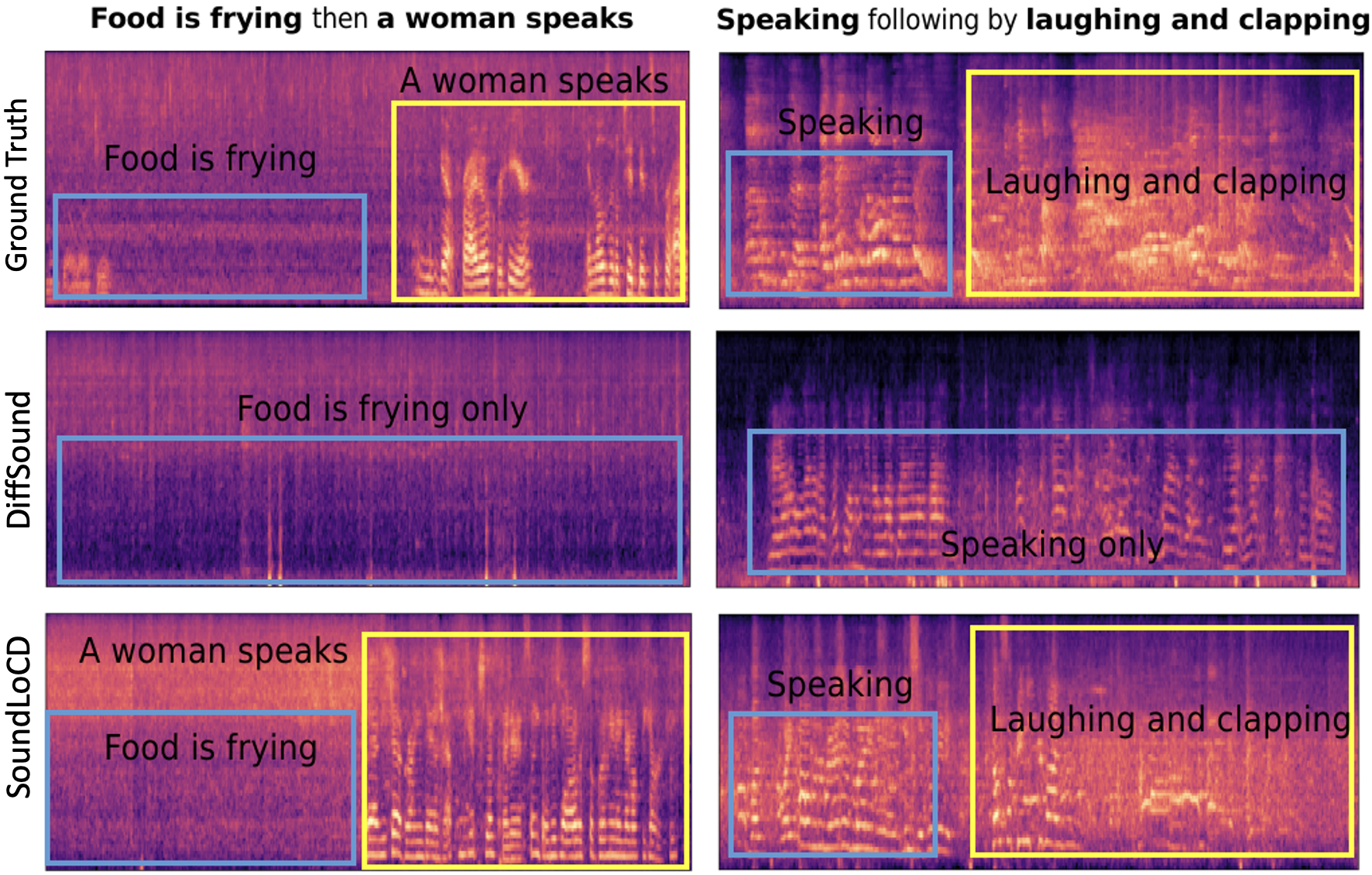}
    \caption{The visualization of generated samples by the DiffSound and SoundLoCD compared with ground truth.} 
    \label{fig:vis}
    \vskip -0.15in
\end{figure}
\section{Experiments and Results}
\noindent\textbf{Dataset.}
We evaluate our approach on AudioCaps~\cite{audiocaps} and ESC50~\cite{piczak2015dataset}. AudioCaps ($\approx$50K audio clips)
using 49256, 494, and 957 audio clips for training, testing, and validation respectively. Each of the audio clips in the training set contains a human-annotated caption while audio clips in the testing set and validation set contain five captions. We use the training set to train our model and verify the model with the validation set. ESC50 contains 2000 environmental sound clips for 10 classes prearranged into 5 folders. We use the first four folders for training and the $5$th folder for testing.

\noindent\textbf{Implementation details.}
We extract the log Mel-spectrogram on a 22050 Hz sampling rate with 1024 frame size, 25\% overlapping, and 80 Mel filter bins. All the audio clips are padded to 10s. The SoundLoCD is trained by batch size 24 and 5e$^{-5}$ contrastive loss weight ($\lambda$) with 10 shuffled sample sizes ($N$) with a maximum of 200 epochs. All the experiments are performed on two NVIDIA GeForce RTX 3090 GPUs. We use a pre-trained MelGAN~\cite{melgan} as the vocoder to obtain waveforms from generated spectrograms.

\noindent\textbf{Evaluation method.}
Following~\cite{diffsound}, we use a series of objective evaluation metrics to verify our method including \textit{Fréchet inception distance (FID), inception score (ISc), Kullbac-Leibler divergence (KL)}, and \textit{kernel inception distance (KID)}. The FID is commonly used to verify the similarity and consistency between generated samples and real samples. While KL measures dissimilarity between the two probability distributions of generated samples and real samples. ISc and KID evaluate sample quality and diversity. 

\subsection{Results and Analysis}\label{exp:LoRA}

In our experiments, SoundLoCD is trained based on a pre-trained DiffSound\footnote{\url{https://github.com/yangdongchao/Text-to-sound-Synthesis}} 
with codebook size 256 and 100 diffusion steps on AudioCaps only.
As we claimed in Section~\ref{sec:intro}, other LDM-based models~\cite{AudioLDM,LLM,Text-Driven,makeaudio,make-audio2} target audio prompting in which training depends on very large-scale dataset, leading to unfair direct performance comparison. Thus, we select DiffSound as the baseline for performance comparison.

We inject LoRA parameters on the whole attention blocks ($W_q,W_k,W_v, W_p$) of the transformer with $r=8$ and $\alpha = 16$. We observed that SoundLoCD achieves better overall performance (see Table~\ref{tab:overal}) on both general sound effects synthesis fidelity, diversity, and text-generated sound correspondences (ISc, KID, and FID). Meanwhile, SoundLoCD significantly reduces the computational requirements for training as it only contains 2.38M trainable parameters. Visualizations in Figure~\ref{fig:vis} indicate that samples generated by SoundLoCD are more related to the text conditions than DiffSound. We then fine-tuned the DiffSound and SoundLoCD on ESC50 in Table~\ref{tab:esc50} which further shows the superiority of the SoundLoCD.

We then study whether text encoders affect the performance of SoundLoCD. We select three typical text encoders from pre-trained models of CLIP~\cite{CLIP}, CLAP~\cite{CLAP}, and T5-small~\cite{T5}\footnote{ Other large pre-trained T5 models are not performing in a 512 dimension, which cannot be directly adapted into the SoundLoCD.}. CLIP is a contrastive language-image pretraining model, T5 studies
text-to-text transformer and CLAP is a contrastive language-audio pretraining model. All of these text encoders embed text tokens into a feature space of dimension 512.
Our experiments indicate that SoundLoCD gets better performance with the CLIP text encoder which is not surprising as the DiffSound checkpoint is trained with CLIP text encoder. CLAP text encoder is trained at an audio waveform level, which might work for continuous latent spaces. While CLIP text encoder is trained at the intrinsically discrete image level may be more suited for discrete latent spaces.

In Table~\ref{tab:LORA_cofg}, we inject LoRA parameters in different types of weights in attention blocks and $r$ where the range of $r$ is selected referring to~\cite{LoRA}. Injecting LoRA weights into the whole attention block reaches the best performance when fixing the $r$. This is consistent with the results of~\cite{LoRA} that it is preferable to adapt LoRA on more weight matrices. The rank size is another critical factor on SoundLoCD, a smaller rank size (e.g., $r=4$) cannot capture enough information in $\Delta W$. However, the model performance will not keep increasing with the increased rank size. We choose the rank size in SoundLoCD to achieve a trade-off between model performance and trainable parameters of the model.

\subsection{Ablation Study} 
Our ablation study (Table~\ref{tab:ablation}) shows that directly applying LoRA on DiffSound with $r=8$ to fine-tune the
whole attention block fails to improve model performance, although the trainable parameters are significantly reduced.
This indicates that DiffSound with LoRA may suffer from over-fitting.
In SoundLoCD, LoRA parameters are optimized by a contrastive learning strategy among $N+1$ parallel diffusion processes which further enhances the connection between text conditions and generating sample beyond the DiffSound, leading to improved performance compared with the baseline.

\begin{table}

\renewcommand{\arraystretch}{0.8}
\renewcommand{\tabcolsep}{2.5mm}
\vskip -0.2in
\caption{Ablation study on CDCD and LoRA in SoundLoCD.}
\begin{center}
\begin{small}
\begin{tabular}{cccccc} 
 \toprule
  & LoRA & CDCD & FID $\downarrow$  & ISc $\uparrow$ \\ 
 \midrule
 DiffSound &  &   & 12.45 &  16.95 $\pm$ 0.87 \\
 DiffSound & \checkmark & & 15.31  & 13.44 $\pm$ 0.73 \\
 SoundLoCD &  \checkmark & \checkmark & \textbf{12.27} &  \textbf{17.36 $\pm$ 1.74}\\ 
  \hline
\end{tabular}
\vskip -0.35in
\label{tab:ablation}
\end{small}
\end{center}
\end{table}

\section{Conclusion}
We proposed SoundLoCD which efficiently reduces the model training cost and further enhances the connections between text conditions and generated samples by a contrastive discrete conditional diffusion model with LoRA, leading to better synthesis results.
Compared to other discrete latent diffusion T2S models, it significantly reduces the training cost while achieving better performance on general fidelity. We studied how different text encoders and LoRA configurations affect SoundLoCD performance and conducted an ablation study to verify the contribution of each part of SoundLoCD. SoundLoCD could support small-scale training of T2S models for creative arts applications and lead to future studies that leverage fine-tuning techniques on large-scale models to improve training efficiency with limited resources.

\bibliographystyle{IEEEbib}
\bibliography{strings,refs}

\end{document}